\documentclass[conference,onecolumn]{IEEEtran}
\usepackage{cite}
\usepackage{graphicx}
\graphicspath{.}
\DeclareGraphicsExtensions{.pdf}

\usepackage{booktabs}
\usepackage[formats]{listings}
\usepackage[english]{babel}
\usepackage[T1]{fontenc}
\usepackage[numbers]{natbib}
\usepackage{multirow}
\usepackage{indentfirst}
\usepackage{amsmath,amssymb,amsfonts}
\usepackage{comment}
\usepackage{graphicx}
\usepackage{listings}
\usepackage{color}
\usepackage{xcolor}

\begin{document}
\title{Anonymity and Confidentiality in \\Secure Distributed Simulation\footnotemark}

\author{\IEEEauthorblockN{Antonio Magnani \quad Gabriele D'Angelo \quad Stefano Ferretti \quad Moreno Marzolla}
\IEEEauthorblockA{Department of Computer Science and Engineering, University of Bologna, Italy\\
Email: \{antonio.magnani, g.dangelo, s.ferretti, moreno.marzolla\}@unibo.it}
}

%\IEEEpubid{978-1-5386-4028-9/17/\$31.00~\copyright2017 IEEE}

\maketitle

\footnotetext{The publisher version of this paper is available at \url{https://doi.org/10.1109/DISTRA.2018.8600922}.
\textbf{{\color{red}Please cite this paper as: ``Antonio Magnani, Gabriele D'Angelo, Stefano Ferretti, Moreno Marzolla. Anonymity and Confidentiality in Secure Distributed Simulation. Proceedings of the IEEE/ACM International Symposium on Distributed Simulation and Real Time Applications (DS-RT 2018)''.}}}

\begin{abstract}
Research on data confidentiality, integrity and availability is gaining momentum in the ICT community, due to the intrinsically insecure nature of the Internet. While many distributed systems and services are now based on secure communication protocols to avoid eavesdropping and protect confidentiality, the techniques usually employed in distributed simulations do not consider these issues at all. This is probably due to the fact that many real-world simulators rely on monolithic, offline approaches and therefore the issues above do not apply. However, the complexity of the systems to be simulated, and the rise of distributed and cloud based simulation, now impose the adoption of secure simulation architectures. This paper presents a solution to ensure both anonymity and confidentiality in distributed simulations. A performance evaluation based on an anonymized distributed simulator is used for quantifying the performance penalty for being anonymous. The obtained results show that this is a viable solution.
\end{abstract}


% -*- LaTeX -*-
\section{Introduction}\label{sec:introduction}

The complexity of many systems that are studied today requires
scalable and efficient modeling and simulation tools. Whether we are
dealing with the Internet of Things (IoT), military scenarios, or
bio-medical systems, these models require huge computation times in
order to obtain credible results~\cite{hpcs16}.  Multilevel, hybrid
and distributed simulation are all helpful techniques that can be
employed to build scalable simulations by aggregating a pool of
resources~\cite{gda-simpat-iot}. Of course, this flexibility has a
price in terms of computation and communication overhead.

Traditionally, the research community working on distributed
simulation focused on increasing the efficiency of Parallel And
Distributed Simulation (PADS) engines to improve scalability. However,
efficiency and scalability are not the only properties that matter in
this context.  In fact, confidentiality and anonymity can pay an even
more important role in some contexts. For example, let us suppose that
we are running a distributed simulation -- say, a highly classified
warfare simulation -- over a pool of servers connected through the
Internet. How can we make sure that the data exchanged among the hosts
can not be intercepted or tampered with?  Sniffing the interactions
among the components of a distributed simulation can provide a lot of
details on the structure of the simulated system. While this problem
is obvious and can be solved using standard end-to-end encryption
techniques (e.g.,~virtual private networks), there are particular
situations where this is not enough. Indeed, encryption can hide the
content of a message so that only the intended recipient is allowed to
recover it; however, encryption alone may still leak the sender and
receiver of a message in the form of their~IP addressed and other
sensitive information.

% ANONYMITY
There are a number of scenarios where anonymity is important. For
example, ``what-if analysis'' is an important support tool for
military decision processes, and it requires both confidentiality and
large-scale simulations to assess multiple war scenarios quickly;
indeed, this kind of simulations must often be executed faster than
real-time. Cyber-warfare is another prominent example that is getting
a lot of attention.  Cyber-warfare is about the actions by a
nation-state (or another international organization) to attack and
possibly damage the computing infrastructures or the communication
network of an opponent. This is usually done by means of computer
viruses/worms, distributed denial-of-service attacks or exploiting
software vulnerabilities. In this scenario, the simulation assets that
are used for managing critical infrastructures or planning the
conventional military operations can be considered both sensitive and
critical~\cite{1530673}.

Of course, the trivial ``solution'' of centralizing all simulation
components so that their interactions do not happen over the public
Internet may not be acceptable. Not only centralized solutions are
more difficult to scale, but in some cases large simulations are
realized by federating components who are owned or controlled by
different organizations that may not be willing to run their code out
of their premises. Many modern simulation models are composed of
components that are run on private or public computing
infrastructures. As a matter of fact, many component-based simulations
are distributed simulations. In general, anonymity can be a useful
add-on in all those scenarios where the simulation working nodes are
deployed on a network. The current trend is to exploit the
computational capabilities offered by cloud computing. The possibility
to take advantage of services which allow a dynamic and rapid
re-configuration of the computation capacity, by adding and removing
working nodes on demand, is fundamental for building simulation models
with a certain degree of complexity. However, there is a number of
threats concerned with the use of third party computation
data-centers, whose interaction requires the use of Internet-based
communication. Moreover, one of the top public cloud computing threats
is concerned with malicious insiders. Even when a data center is
trusted, we still might want to avoid tracking where the simulation
components of a distributed simulation are executed.

% CONFIDENTIALITY AND AVAILABILITY
Confidentiality and availability are other fundamental requirements
for distributed simulation when dealing with sensitive data and
applications. While simulation can be used to model and understand
cyber-threats and cyber-warfare, strategies are needed to prevent that
through these same threats unauthorized entities gain access to the
simulated data or alter the simulation behavior. More specifically,
availability is a main requirement in every distributed application
and system. It is worth noting that availability and anonymity are, at
least in part, correlated. In fact, being anonymous \emph{can} prevent
specific types of attacks by some malicious subjects (e.g.~distributed
denial of service attacks), since the attackers cannot know where
simulation components are located and how to reach them.

In this work we discuss how to integrate the anonymity functionalities
provided by a free software to enable anonymous communication
(i.e.,~Tor) in the ART\`IS PADS
middleware~\cite{pads,gda-simpat-2017}. This involves both modifying
the design of the simulation middleware (in order to make it able
communicating using Tor) but also preventing information leaks that
could disclose sensible data of a simulation component to other parts
of the distributed simulation. We evaluate the performance of the
distributed anonymized simulation to quantify the overhead introduced
by the Tor layer on the simulation execution time. Our results
demonstrate that it is possible to build a distributed simulation that
supports a high degree of anonymity of its components. It is worth
noting that the studied approach does not solve all aspects related to
the security of distributed simulations. In fact, being anonymous
requires the adoption of operational security practices that are not
limited to the mere installation (or usage) of a software. It is also
important to underline the price to pay in terms of performance for
being anonymous. In this work, the cost of anonymity in distributed
simulation is investigated and quantified.

The remainder of this paper is organized as
follows. Section~\ref{sec:related-work} describes the background and
related works. In Section~\ref{sec:simulator} we present the anonymous
distributed simulator. Results from a performance evaluation are
discussed in Section~\ref{sec:perfeval}. Finally,
Section~\ref{sec:conclusions} provides some concluding remarks.
 
% -*- LaTeX -*-
\section{Background and Related Work}\label{sec:related-work}

In this section we provide the background information that is required
to understand the rest of the paper, including related works on
simulation and security.

\subsection{Background}

\subsubsection{Tor}
is an open network protocol~\cite{tor}, for which free implementations
are provided, that is designed and operated for improving privacy and
security on the Internet. The aim of~Tor is to make anonymous
communications possible. This is achieved through an overlay network
over the public Internet that is currently composed of about~$6000$
volunteer-operated servers. The basic component of Tor is the
\textit{onion routing algorithm} that allows end-to-end anonymity
through multiple layers of encryption. In Tor, the user data (in our
case, the simulation events) is encrypted multiple times and sent
through a virtual circuit composed of a set of~Tor relays
(i.e.,~application-level routers, also called \emph{onion routers})
that are randomly selected. A virtual circuit has a limited life span,
meaning that after a certain amount of time a virtual circuit is no
longer valid and must be rebuilt. It is worth noting that Tor encrypts
both the user data and some metadata such as the~IP address of the
next destination of a packet. Each Tor relay can decrypt one layer of
encryption only, so it can only know the next relay in the virtual
circuit to forward the data to. In presence of correct operational
security, and assuming that there are no vulnerabilities in all
software components of the Tor network (including the communicating
parties), this approach can hide the users' location from network
surveillance and traffic analysis. Tor has been initially conceived
for providing anonymity to users that access online services, but it
can also be used to hide the true identity of servers. The servers
that are reachable through the Tor network are called \textit{hidden
  services}. The~IP address of a service would obviously reveal its
identity and location. For this reason, each hidden service is
accessible only using its \textit{onion address}, an identifier
provided by the Tor network. An onion address allows clients to access
a service without knowing its~IP address. All client connections to
hidden services are encrypted end-to-end, and therefore eavesdropping
is not an effective attack method. A detailed description of~Tor is
outside the scope of this paper; a comprehensive introduction can be
found in~\cite{6266172}.

\subsubsection{Parallel And Distributed Simulation}
the traditional approach to the implementation of a simulation model
is through a serial program. This means that a single execution unit
is in charge of running the whole simulation model. In recent years,
sequential simulations have shown many
limitations~\cite{gda-simpat-2017} mainly in terms of lack of
scalability when dealing with large and complex models. A Parallel and
Distributed Simulation (PADS) takes advantage of multiple execution
units to efficiently handle large simulations~\cite{Fuj00}. These
execution units can be distributed across the Internet, or grouped as
massively parallel computers or multicore processors. In the~PADS
approach, the simulation model is partitioned in submodels, called
Logical Processes (LPs) which can be evaluated concurrently by
different execution units. More precisely, the simulation model is
described in terms of multiple interacting entities that are assigned
to different LPs. Each~LP runs on a different execution unit, where an
execution unit acts as a container of a set of entities. The
simulation progress through the exchange of timestamped messages,
representing events that must be exchanged among entities.

\subsection{Related Work}

Even if security is often listed among the requirements for creating
reliable and scalable distributed simulations for real-time decision
making~\cite{Fujimoto:2016:RCP:2892241.2866577}, the research done on
this topic is quite limited. In fact, more attention has been devoted
to the usage of simulation for investigating systems security, that is
clearly a totally different topic with respect to the problem
addressed in this paper.

A large part of the research on the security of distributed
simulations is strictly related to the High Level Architecture
(HLA). The~HLA is a general architecture that aims at improving the
interoperability of distributed simulators~\cite{HLA} through the
composition of software components called \emph{federates}.  In
essence, a federate represents a portion of a (possibly large)
simulation model. The main security issues of~HLA are discussed
in~\cite{Elkins:2001:SIH:564124.564239}. The authors investigate the
usage of the IPsec protocol to provide confidentiality to the
federates communicating through the HLA runtime infrastructure
(RTI). Additionally, the authors envision the adoption of a public key
infrastructure for access control on specific parts of the simulation
model (e.g.,~the federation object model).

In~\cite{oatao2056} and~\cite{Bieber98securityextensions}, the authors
discuss the security of distributed simulation and describe the
trusted third party architecture that is implemented in the
CERTI/ONERA RTI. The threat model considered by the authors assumes
that each federate can trust some components of the~RTI, but does not
trust the other federate components. The proposed approach relies on
secure domains and on the use of Generic Security Services Application
Program Interface to secure communications between remote federates
and the~RTI.

More recently, in~\cite{5975588} the authors propose the usage of
secure protocols (e.g.,~HTTPS) and an authentication framework based
on a public key infrastructure to enable the execution of a
specific~HLA RTI on a cloud environment. In this case, the main goal
is to provide data confidentiality and integrity of services.

Among the few research efforts that are not strictly related to~HLA,
in~\cite{Mills-Tettey:2003:SIA:786111.786229} the authors discuss the
challenges in developing a security framework for the Agent-Based
Environment for Linking Simulations (ABELS) framework. Many security
aspects such as authentication, authorization and integrity of the
distributed simulator are considered. Specifically, the proposed
system is based on different security modes and a brokering security
mechanism.

Finally, in~\cite{Norling2007SecureDC} the same security aspects are
discussed when applied to a distributed co-simulation in which
different simulation models are interconnected together.

% -*- LaTeX -*-
\section{AnonSim: an Anonymous Simulator}\label{sec:simulator}

This section presents AnonSim, a new distributed simulator that
implements a method for anonymizing the communication streams among
LPs. AnonSim is based on the GAIA/ART\`IS distributed simulation
middleware~\cite{gda-perf-2005} and relies on~Tor for the exchange of
information among LPs. We first provide some information on the
GAIA/ART\`IS middleware, and then describe the modifications that have
been applied to achieve anonymity.

In some systems it is possible to find that some LPs need to be
anonymous while this might not be true for others. For the sake of
simplicity, in this paper we assume that the anonymized version of the
simulator is applied to all the LPs that are part of the distributed
simulation (i.e.,~all LPs use Tor). Under the implementation
viewpoint, this means that all the communications between the LPs must
be done relying on the usage of the \textit{onion address} of the LPs.

\subsection{GAIA/ART\`IS logical architecture}

ART\`IS~\cite{gda-perf-2005} is a distributed middleware that allows
seamless execution of sequential/parallel/distributed simulation
models using different communication (e.g.,~shared memory, TCP/IP,
MPI) and synchronization methods (e.g.,~time-stepped, conservative,
optimistic). The GAIA framework~\cite{gda-simpat-2017} runs on top of
ART\`IS and provides a high level application program interfaces that
greatly simplifies the development of simulation models according to
an agent-based approach. Furthermore, GAIA implements strategies for
balancing the communication and computation load among the LPs based
on adaptive partitioning, i.e.,~dynamic self-clustering of the
simulation agents. This can reduce the communication overhead and
achieve a more uniform usage of the computation resources, therefore
speeding up the simulation.

\subsection{AnonSim communication architecture}

An important component of ART\`IS is the SImulation MAnager
(SIMA). The SIMA is used only for bootstrapping the simulation runs
and, in this specific phase, it works as the central coordinator of
the LPs.  Specifically, it records information about all the LPs
involved in the simulation, and at startup, the SIMA performs the role
of simulation initiator. Thus, at the simulation bootstrap each LP
must contacts the SIMA to receive information about the other
participants in the simulation run, and the simulation starts only
once the transmission channels between the LPs have been initialized.

\begin{figure}[h]
\centering
\includegraphics[width=0.60\linewidth]{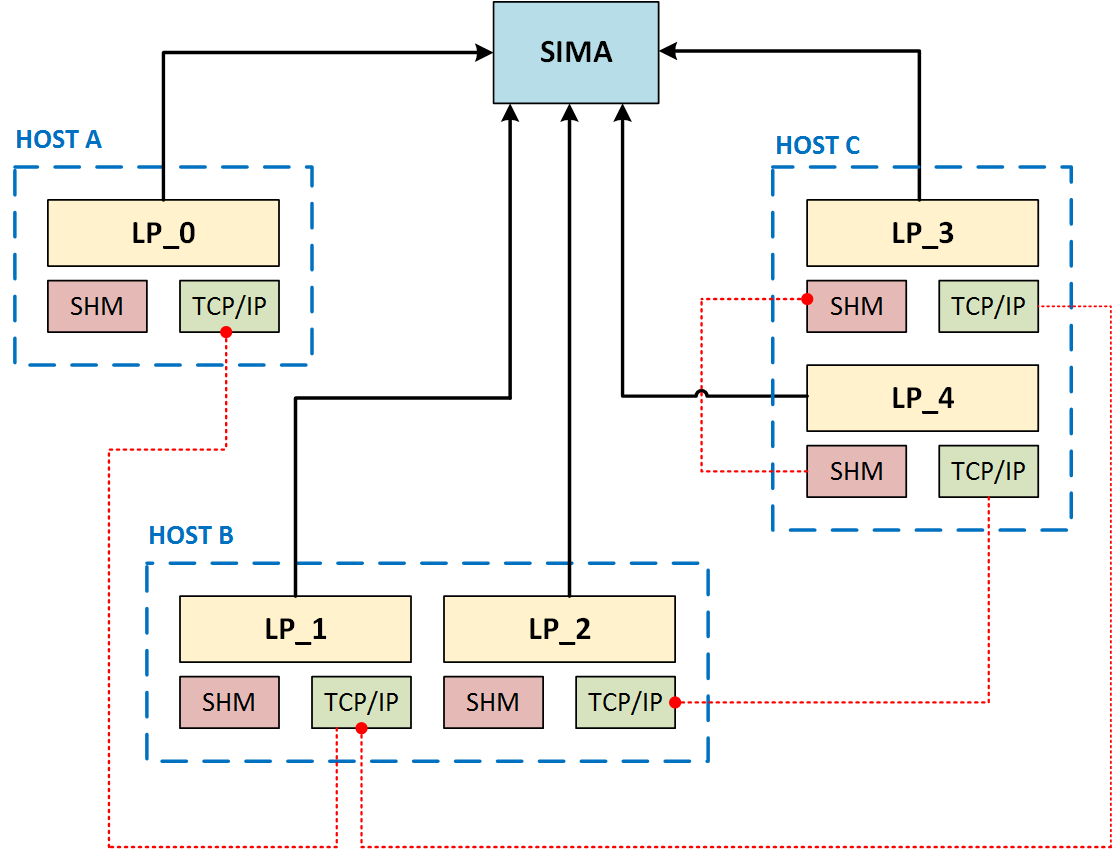}
\caption[Communication architecture of ART\`IS]{Communication architecture of ART\`IS.\label{fig:artis-comm}}
\end{figure}

Figure~\ref{fig:artis-comm} depicts the communication architecture of
ART\`IS. Each~LP contacts the~SIMA to register itself (black lines in
the figure). Since the~SIMA knows in advance the number of LPs in each
simulation run, it waits until every~LP has registered.  The
simulation starts when each~LP receives from the SIMA the information
about all other~LPs. After that, the nodes can start interacting by
exchanging messages (red dotted lines). In the figure it is shown that
LPs can communicate through two different mechanisms: shared memory
(SHM in the figure), or (secured) TCP connections.  Shared memory is
used only among LPs that are executed on the same physical host.

In AnonSim, the SIMA receives only anonymized information about all
the LPs, meaning that compromising the SIMA is not sufficient, e.g.,
for discovering the real location of the~LPs. All TCP communications
are tunneled through Tor by using the Socket Secure (SOCKS)
protocol. Thus, each~LP (and the SIMA) relies on SOCKS to interoperate
with the simulation middleware and network communications, and this
imposes that each host running a LP has a running local SOCKS server;
in practice SOCKS provides encryption and anonymity for transmitted
messages. All the simulation components are in fact Tor hidden
services. As described above, the hidden service functionality is
essential for providing anonymization of the simulation
components. The interaction of SOCKS with the Tor protocol can be
summarized as follows:
\begin{enumerate}
\item Each LP connects to its local SOCKS server, instead of directly
  connecting to the destination hidden service (e.g.,~SIMA or other
  LPs);
\item The local SOCKS server sends a \textit{connect request} to the
  receiver using onion address of the destination;
\item The hidden service (SIMA or LP) replies to the SOCKS server with
  an \textit{answer code}: if the request is accepted by the
  destination then the SOCKS server tries to establish a connection to
  the destination hidden service.
\end{enumerate}

\begin{figure*}[htbp]
\centering%
\includegraphics[width=0.60\linewidth]{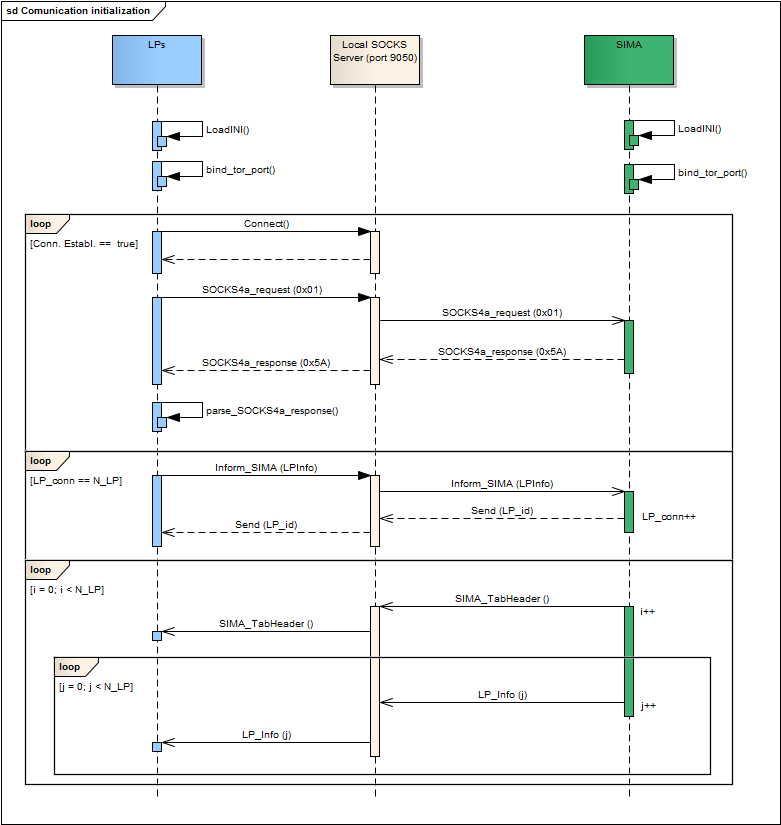}
\caption{Sequence diagram of the communication initialization using ART\`IS APIs. The interaction between the LP, the SIMA and the local SOCKS server are reported.}\label{fig:SeqDiag}
\end{figure*}

Using this communication mechanism, each LP first registers to the
SIMA in order to know how to contact other LPs. Due to the nature of
the Tor network, LPs may have to retry several times the initial SOCKS
connect request. Once the SOCKS response is received, an LP has its
own connection to communicate with the SIMA. In its first message, the
LP provides to the SIMA all the needed information for being
contacted, such as its onion address and TCP port. Then the SIMA
answers with a unique identifier that is specific for this LP. Once
all information about each LPs have been collected, the SIMA
broadcasts is formation of all LPs in the simulation run. Note that
all the information above is completely anonymous. In fact, each LP is
represented only by the identifier assigned by the SIMA (which acts as
a pseudonym) and is reachable only using the onion address of its
hidden service that is provided by Tor. Figure~\ref{fig:SeqDiag}
summarizes the interactions during this initialization phase.

After the initialization, each LP establishes a connection to all
other LPs, thus creating a complete anonymous undirected overlay
graph. In particular, since each network connection is bidirectional,
a LP instantiates a connection only with those LPs that have a lower
identifier (the one assigned by the SIMA). This avoids the creation of
redundant pairs of (bidirectional) network connections.  After this
step, the simulation can start. All the communications will pass
through these SOCKS connections, which are kept anonymous by the Tor
network.

AnonSim is available for peer review and it will be integrated in the
next release of the GAIA/ART\`IS simulation middleware~\cite{pads}.

% -*- LaTeX -*-
\section{Performance Evaluation}\label{sec:perfeval}

In this section we evaluate the performance of AnonSim to assess the
impact of the network anonymization provided by Tor.

\subsection{Simulation model}

The model used in this performance evaluation is the
``MIGRATION-WIRELESS'' discrete event simulation model provided in the
ART\`IS package~\cite{pads,gda-simpat-2017}. The model is agent-based
and consists of a set of Simulated Mobile Hosts (SMHs) wandering over
a bi-dimensional toroidal space according to the Random Waypoint (RWP)
mobility model~\cite{rwp,gda-simpat-2017}. The SMHs interact by
exchanging ping messages that are sent by each SMH to all SMHs that
are within a fixed distance. From the simulator point of view, each
SMH is implemented as a separate entity; the SMHs are initially
randomly distributed across the LPs. The model uses the time-stepped
synchronization scheme~\cite{gda-simpat-2017}.

\subsection{Simulation architecture}

Given the expected use cases for AnonSim (i.e., running a distributed
simulation over the Internet with confidentiality and anonymity
requirements), the ``MIGRATION-WIRELESS'' model has been executed on
three interacting hosts that are physically distributed through
Europe.  More specifically, each host will accommodate a single
LP. The computational resources were supplied by two cloud
providers. In particular, two instances were hosted by Amazon
EC2~\cite{EC2} and one by Okeanos~\cite{okeanos}. The two~EC2 machines
were instances of the type \textit{ec2.micro} located in Ireland
(named \textit{EC2.dublin}) and in Germany (named
\textit{EC2.frankfurt}). The remaining \textit{Okeanos} host was
physically placed in Greece. The detailed characteristics of the cloud
instances used in this setup were:
\begin{itemize}
\item \textit{ec2-instaces}: 1 vCPUs, 2.5 GHz, Intel Xeon Family, 1
  GiB memory; network performances declared: low to moderate.
\item \textit{okeanos}: 4 vCPUs, 2.1 GHz, AMD-V Family, 4 GiB memory;
  network performances not specified.
\end{itemize}

\begin{figure}[t]
\centering
\includegraphics[width=0.60\linewidth]{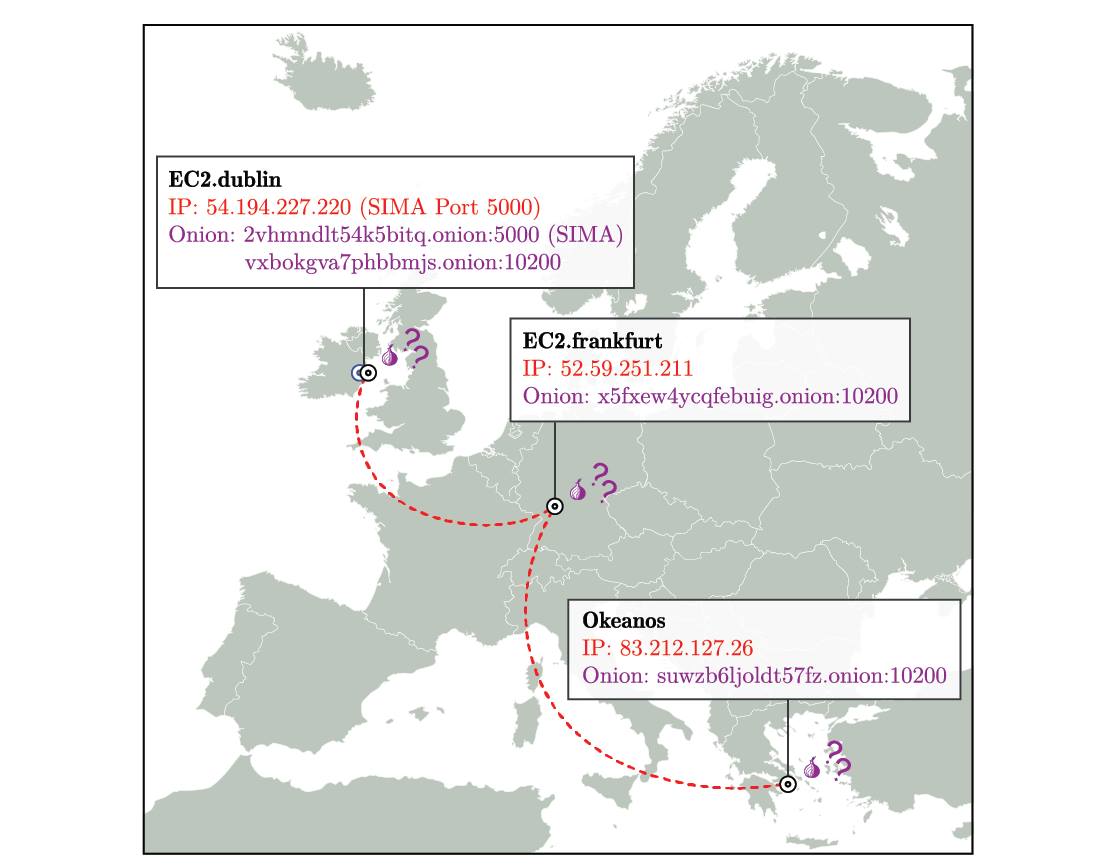}
\caption{Simulation distribution of the hosts used in performance evaluation of AnonSim.}
\label{fig:LPComm}
\end{figure}

Figure~\ref{fig:LPComm} shows the geographical location of the hosts
running the anonymized distributed simulation. For each host, the~IP
address (red) and the onion address (purple) are also shown. The~SIMA
was placed in the \textit{EC2.dublin} instance. For this reason, this
host has two different onion addresses: one for the SIMA and the other
one for the LP.

It is worth noting that it is easy to record the path followed by the
TCP segments used for delivering the simulation messages when the
distributed simulation is run without using the anonymization provided
by Tor, e.g., using applications like \textit{traceroute}. On the
other hand, in presence of the Tor network the path followed by the
encapsulated TCP segments is harder to trace due to the use of virtual
circuits and the characteristics of the Tor anonymization protocol.

\subsection{Tor performances}

The use of the Tor overlay network instead of the physical
communication layer leads to a higher communication latency;
therefore, an overall increase of the Round-Trip Time (\textit{RTT})
is to be expected. We use the \textit{tcpping}~\cite{tcpping} software
to get an estimate of the RTT.  \textit{tcpping} mimics the behavior
of the classic \emph{ping} application, but at the TCP layer. The use
of \emph{tcpping} is necessary since Tor does not support ICMP packets
employed by the \emph{ping} program.

\begin{table}[h]
  \begin{center}
    \caption[RTTs of standard TCP/IP and Tor communications]{Round-Trip Times of standard TCP/IP and Tor communications: the results are obtained with $200$ different \textit{ping}, one every $3$ sec., between our cloud instances.\\ $\overline{x}$ are the mean values; $\sigma$ are the standard deviations. }
\begin{tabular}{@{\extracolsep{.5em}}lcccc}
\toprule
& \multicolumn{2}{c}{\textbf{TCP/IP}} &\multicolumn{2}{c}{\textbf{Tor}} \\ 
& \multicolumn{1}{c}{$\overline{x}$ (ms)} & \multicolumn{1}{c}{$\sigma$ (ms)} & \multicolumn{1}{c}{$\overline{x}$ (ms)} &  \multicolumn{1}{c}{$\sigma$ (ms)} \\
\cmidrule{2-3} \cmidrule{4-5}
\multicolumn{1}{ c }{\textit{EC2.dublin - Okeanos}} & \multicolumn{1}{ c }{92.91} & 0.75 & \multicolumn{1}{ c }{326.42} & 278.52 \\
\multicolumn{1}{ c }{\textit{Okeanos - EC2.frankfurt}} & \multicolumn{1}{ c }{67.14} & 0.16 & \multicolumn{1}{ c }{282.74} & 104.83 \\ 
\multicolumn{1}{ c }{\textit{EC2.frankfurt - EC2.dublin}} & \multicolumn{1}{ c }{20.84} & 0.20 & \multicolumn{1}{ c }{540.74} & 54.5 \\
\bottomrule
\end{tabular}
\label{Table:ping}
\end{center}
\end{table}

Table \ref{Table:ping} shows the RTTs between the hosts, using plain
TCP (without Tor) and with Tor. As expected, Tor increases the network
latency considerably, and also increases the variability of such
latencies (i.e.,~introduces a wide standard deviation). Such high
variability is due to the ephemeral nature of the virtual circuits
provided by Tor. Such variability can be explained by considering
several factors: the nature of the relays which compose the Tor
network; the positions of the onion routers in a circuit; finally --
and extremely relevant -- the fact the circuits change over time. The
latter implies a costly reconstruction of new circuits.

\begin{figure}[!h]
\centering
\includegraphics[width=0.60\linewidth]{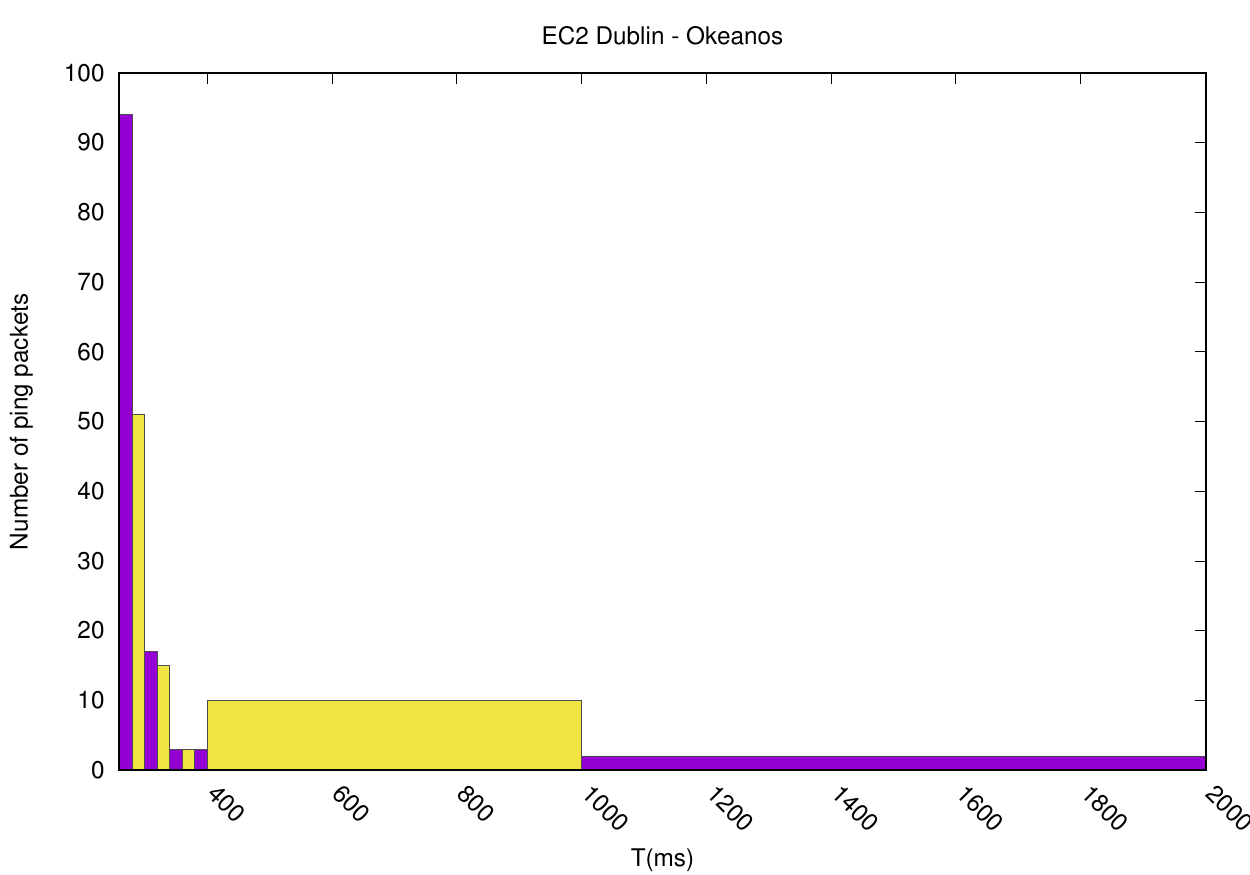}
\caption[RTT distribution of $200$ packets (encapsulated by Tor) between EC2.dublin and Okeanos.]{RTT frequency distribution of 200 ping packets (encapsulated by Tor) between the EC2.dublin instance and the Okeanos instance. The mean value ($\overline{x}$) is \textit{326.42 ms} and the standard deviation ($\sigma$) is \textit{278.52 ms}.}
\label{fig:tor_amazon_to_okeanos}
\end{figure}

\begin{figure}[!h]
\centering
\includegraphics[width=0.60\linewidth]{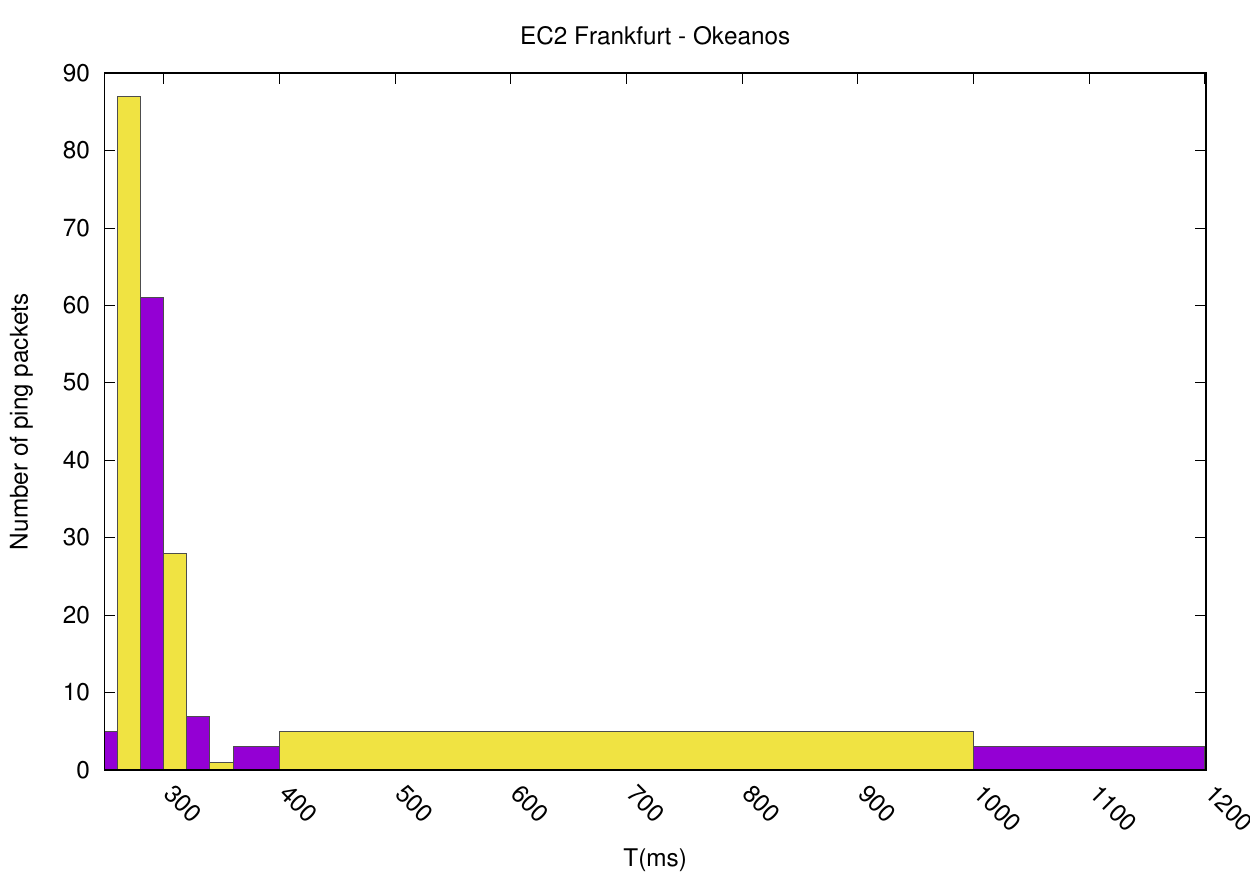}
\caption[RTT distribution of $200$ packets (encapsulated by Tor) between Okeanos and EC2.frankfurt.]{RTT frequency distribution of 200 ping packets (encapsulated by Tor) between the Okeanos instance and the EC2.frankfurt instance. The mean value $\overline{x}$ is \textit{282.75 ms} and the standard deviation $\sigma$ is \textit{104.83 ms}.}
\label{fig:tor_amazon2_to_okeanos}
\end{figure}

\begin{figure}[!h]
\centering
\includegraphics[width=0.60\linewidth]{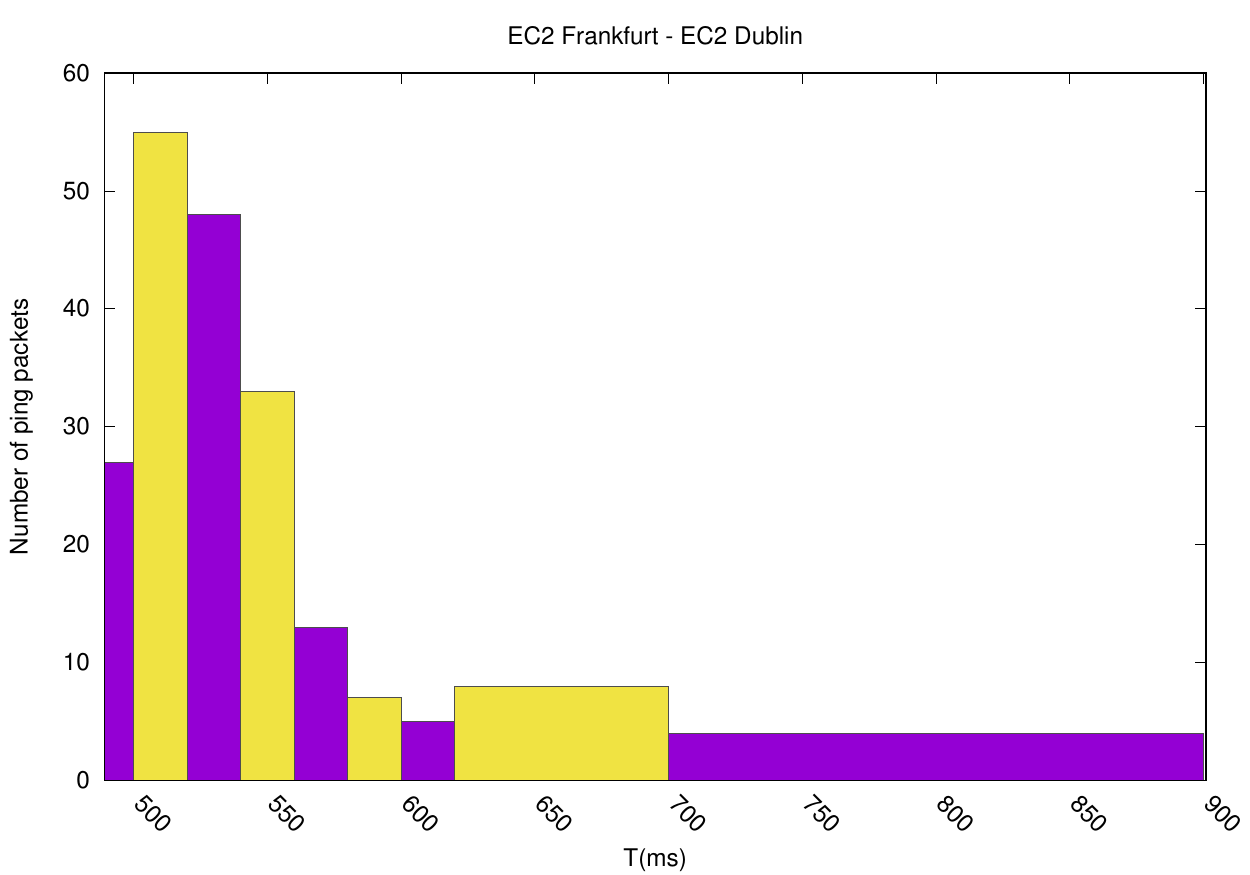}
\caption[RTT distribution of $200$ packets (encapsulated by Tor) between EC2.frankfurt and EC2.dublin.]{RTT frequency distribution of 200 ping packets (encapsulated by Tor) between the EC2.frankfurt instance and the EC2.dublin instance. The mean value $\overline{x}$ is \textit{540.74 ms} and the standard deviation is \textit{54.50 ms}.}
\label{fig:tor_amazon_to_amazon}
\end{figure}

To illustrate the high RTT variability caused by Tor, we show in
Figures~\ref{fig:tor_amazon_to_okeanos},
\ref{fig:tor_amazon2_to_okeanos} and~\ref{fig:tor_amazon_to_amazon}
the RTT distribution of~$200$ ping packets.  We remark that the
distribution has been obtained during a single execution of a
simulation. Due to the nature of the Tor network and the volatility of
its virtual circuits, a considerably larger number of experiment would
be needed to construct a reliable statistics. Despite this, we report
our results to provide some quantitative evidence on the amount of
delay that might be introduced by Tor, and that will affect the
results provided in the remaining part of this section.

\subsection{AnonSim execution results}

The tests that have been performed to assess the performance of
AnonSim are classified as follows:
\begin{itemize}
\item \textbf{\textit{TCP/IP (or Tor) ALL\_OFF}}: the load balancing
  mechanism provided by GAIA (i.e., the dynamic migration of SMHs
  based on self-clustering) is \emph{disabled}. This means that
  every~LP contains the same SMHs for the whole duration of the
  simulation runs.
\item \textbf{\textit{TCP/IP (or Tor) ALL\_ON}}: the load balancing
  mechanism provided by GAIA is \emph{enabled}. The standard
  self-clustering heuristic provided by~GAIA is used; no tuning of the
  heuristic parameters has been performed.
\end{itemize}

The results shown in the following of this section are obtained by
averaging $10$ simulation runs for each scenario.

\begin{table}[h]
  \centering
  \caption{Wall-Clock-Time (WCT) for the executions of the simulation model with 3000 SMHs. All values are in seconds.}
\begin{tabular}{lrrrrr}
\toprule
\textbf{Configuration}   & \textbf{Mean WCT} & \textbf{Std Dev} & \textbf{Min} & \textbf{Max} & \textbf{CI}  \\
\midrule
\textit{TCP/IP ALL\_OFF} & 130 &   1 & 128 &  132 &   1 \\
\textit{TCP/IP ALL\_ON}  & 107 &   4 & 103 &  115 &   2 \\ 
\textit{Tor ALL\_OFF}    & 924 & 414 & 508 & 1684 & 216 \\
\textit{Tor ALL\_ON}     & 529 & 100 & 425 &  709 &  52 \\
\bottomrule
\end{tabular}
\label{Table:3000se}
\end{table}

\begin{figure}[h]
\centering
\includegraphics[width=0.60\linewidth]{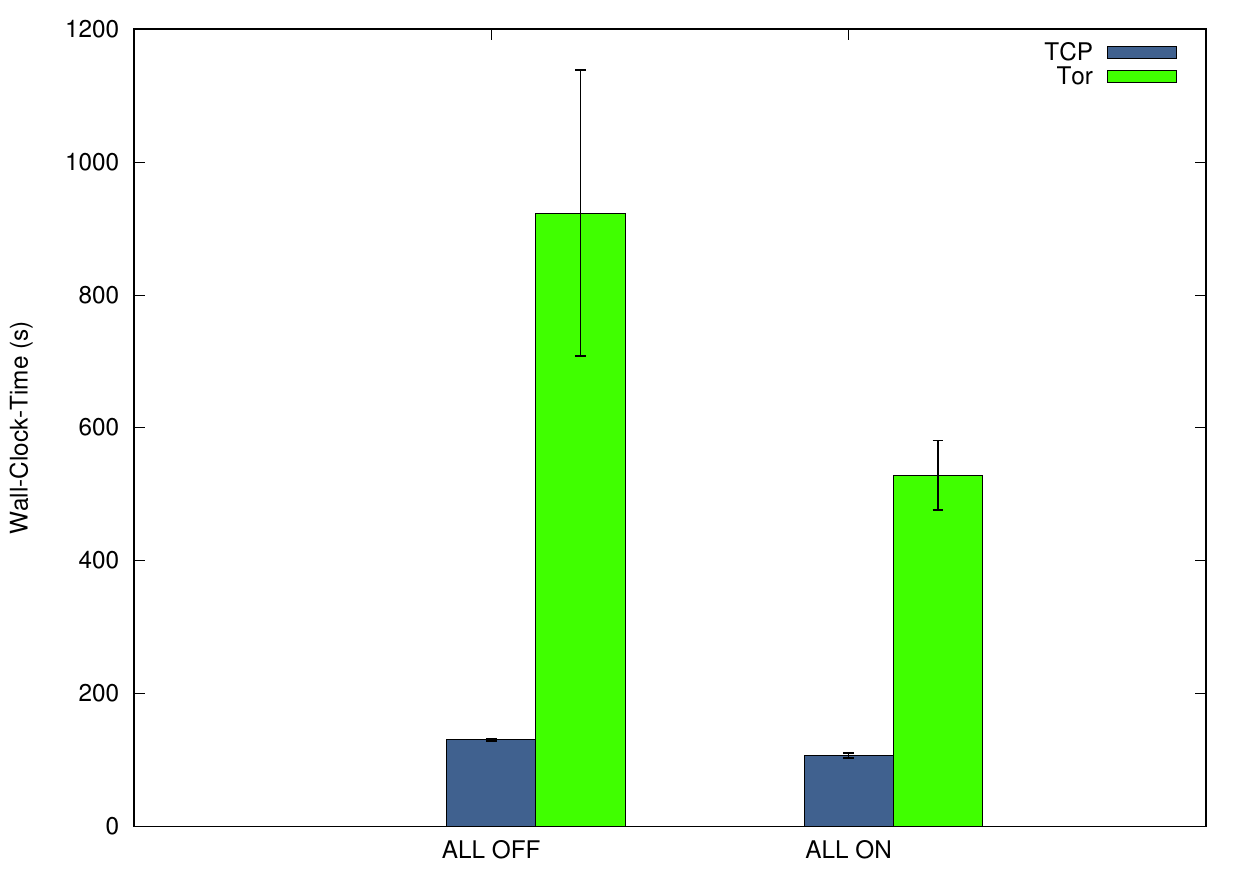}
\caption{Average WCTs for the simulation model with $3000$ SMHs.}
\label{fig:3000se}
\end{figure}

In Table \ref{Table:3000se} we report the average WCT of the
simulation model with $3000$ SMHs, its standard deviation, the minimum
and maximum WCT recorded over the $10$ independent runs, and the
half-width of the confidence interval (CI) for the mean~WCT computed
at $90\%$ confidence level. We observe that, when run with load
balancing turned off, AnonSim requires on average seven times the~WCT
required by the plain setup without Tor. As reported before, this is
expected since the overhead added by the Tor virtual circuits is not
negligible. The wide standard deviation of the~WCT confirms that
during the different sample runs, the Tor circuits changed, altering
the overall performance of the simulator.

When adaptive load balancing is enabled, the speedup of the simulator
using adaptive load balancing with respect to the base version not
using it (computed as the ratio between \{TCP/IP, Tor\} ALL\_OFF and
\{TCP/IP, Tor\} ALL\_ON) is approximately
$1.75$. Figure~\ref{fig:3000se} shows the reduction of the wall-clock
time and, more importantly that is obtained thanks to a significant
reduction of the number of messages exchanged among the LPs, that is
one of the beneficial effects of the dynamic self-clustering provided
by GAIA. While the interaction between Tor and the self-clustering
heuristics is quite difficult to analyze, and outside the scope of
this paper, we can see that the total number of~SMHs migrations (for
self-clustering) requested by GAIA is on average $7,582.40$ when Tor
is enabled and $7,677.00$ when it is not used. In other words, it
seems that Tor does not affects significantly the ability of GAIA to
provide communication and computational load-balancing.

\begin{table}[h]
  \centering
  \caption{Wall-Clock-Time (WCT) measured for the executions of the simulation model with $6000$ SMHs. All values are in seconds.}
\begin{tabular}{lrrrrr}
\toprule
\textbf{Configuration}   & \textbf{Mean WCT} & \textbf{Std Dev} & \textbf{Min} & \textbf{Max} & \textbf{CI}  \\
\midrule
\textit{TCP/IP ALL\_OFF} & 528  &  26 & 466 & 554  &  14 \\
\textit{TCP/IP ALL\_ON}  & 167  &  13 & 157 & 198  &   7 \\ 
\textit{Tor ALL\_OFF}    & 1586 & 826 & 939 & 3403 & 430 \\ 
\textit{Tor ALL\_ON}     & 1022 & 203 & 736 & 1336 & 106 \\
\bottomrule
\end{tabular}

\label{Table:6000se}
\end{table}

\begin{figure}[!h]
\centering
\includegraphics[width=0.60\linewidth]{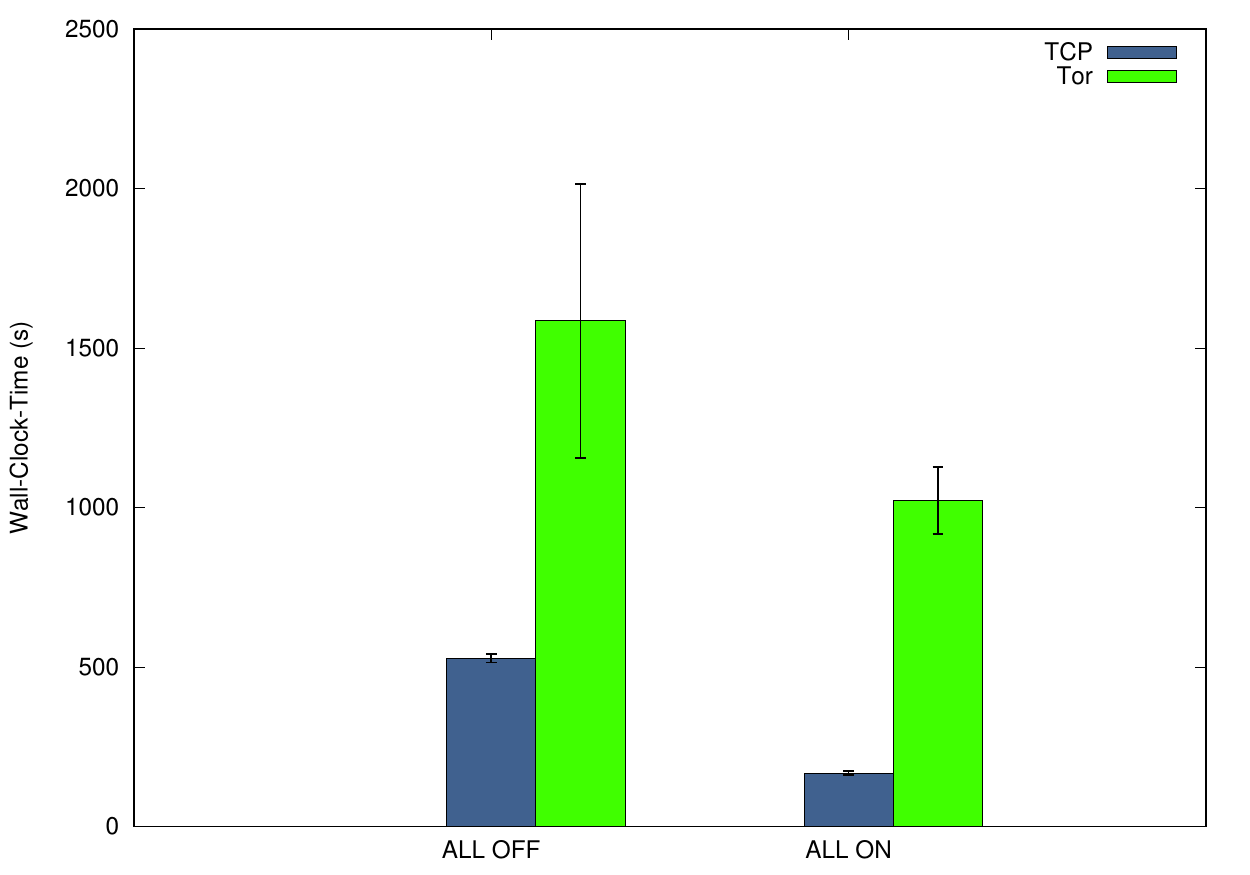}
\caption{Average WCTs for the simulation model with 6000 SMHs and relative confidence intervals.}
\label{fig:6000se}
\end{figure}

As shown in Table~\ref{Table:6000se}, similar results are obtained
when the number of~SMHs is doubled. Figure~\ref{fig:6000se} shows a
WCT that is lower for the TCP/IP execution version with respect to the
Tor one. In this case, the speedup obtained thanks to GAIA is $3.15$
while in the anonymous version the speedup is $1.55$. It is worth
noticing that increasing the number of SMHs causes a higher workload
on the LPs (whose number remained constant with respect to the case
with $3000$ SMHs) both in terms of communication and computation. This
makes it more difficult for GAIA to improve the clustering of SMHs
thus resulting in a higher amount of messages exchanged in the
distributed simulator.

\begin{table}[h]
  \centering
  \caption{Wall-Clock-Time (WCT) measured for the executions of the simulation model with $9000$ SMHs. All values are in seconds.}
\begin{tabular}{lrrrrr}
\toprule
\textbf{Configuration}   & \textbf{Mean WCT} & \textbf{Std Dev} & \textbf{Min} & \textbf{Max} & \textbf{CI}  \\
\midrule
\textit{TCP/IP ALL\_OFF} & 1156  &   58 & 1076 & 1220 &  30 \\ 
\textit{TCP/IP ALL\_ON}  & 397   &   69 &  344 &  505 &  36 \\
\textit{Tor ALL\_OFF}    & 2275  & 1010 & 1144 & 3822 & 526 \\
\textit{Tor ALL\_ON}     & 1720  &  279 & 1232 & 2072 & 145 \\
\bottomrule
\end{tabular}
\label{Table:9000se}
\end{table}

\begin{figure}[!htb]
\centering
\includegraphics[width=0.60\linewidth]{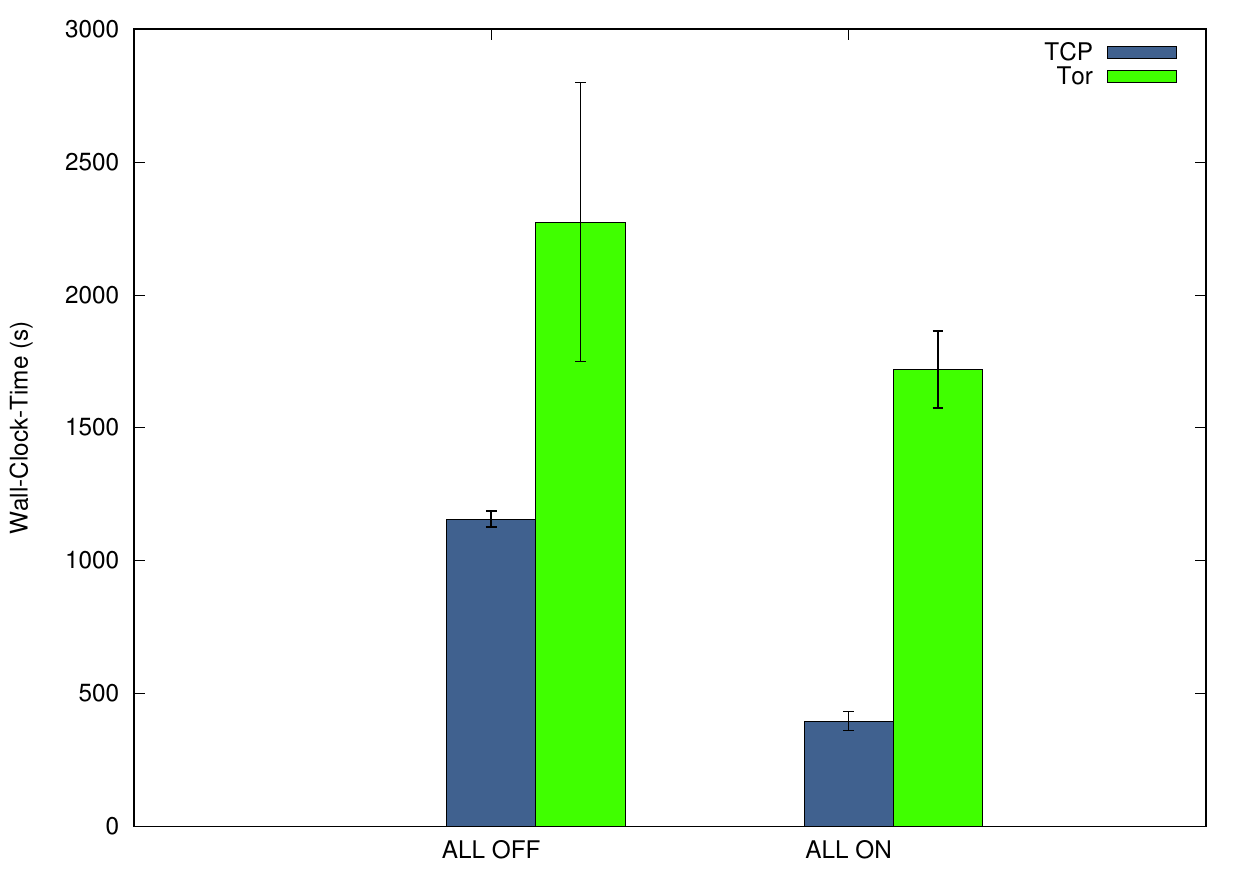}
\caption[WCTs for simulation with 9000 SMHs]{Average WCTs for the simulation model with 9000 SMHs and relative confidence intervals.}
\label{fig:9000se}
\end{figure}

All the considerations reported above are confirmed when the amount of
SMHs is set to 9000 (see Table \ref{Table:9000se} and
Figure~\ref{fig:9000se}). The speedup for the anonymous version is
further reduced to $1.32$ whilst for the TCP/IP version is $2.91$.

\subsection{Reliability}

The main conservative synchronization mechanisms (e.g., the
Chandy-Misra-Bryant NULL messages
protocol)~\cite{Chandy:1981:ADS:358598.358613} and the time-stepped
approach~\cite{fujimoto,gda-marzolla} assume that the communication
channel among the LPs is reliable, i.e., no message is lost or
corrupted. This assumption is mostly true in conventional distributed
simulations relying on the TCP protocol. However, the assumption is no
longer true when we consider TCP connections that are encapsulated in
Tor virtual circuits. Tor virtual circuits are ephemeral for security
reasons; therefore, the periodic restructuring activity of the virtual
circuits can result in transient connection failures that are observed
at the application level. Indeed, during our tests we discovered that
the TCP connections encapsulated in the Tor virtual circuits are
unable to deal with the underlying failure of virtual circuits. This
affects the reliability of the simulator and the visible effect is
that some simulation runs fail due to the unpredictable behavior of
the communication infrastructure.

This issue can be addressed in may ways. Firstly, it is possible to
implement solutions that address the more general problem of
fault-tolerance in distributed simulation. For example,
in~\cite{gda-dsrt-2016} we have proposed an approach based on
functional replication that is capable of tolerating crash-failures of
computing nodes and offers some protection against Byzantine
failures. Secondly, it would be possible to check the correct delivery
of simulation events among the LPs by adding an additional reliability
layer on top of the TCP-based communications. This mechanism should be
implemented in each~LP and it would be in charge of verifying that all
the messages have been correctly received and, in case of failures, of
triggering application layer retransmissions. Thirdly, it would be
possible to port the whole simulator to a synchronization protocol
such as Time Warp~\cite{timewarp} that does not assume the
reliability of communication channels. Of course, this would require a
large scale refactoring of the GAIA/ART\`IS simulator; additionally,
it has been shown that Time Warp may not be the most efficient
protocol for all kinds of simulation models or execution
architectures~\cite{gda-marzolla}.

% -*- LaTeX -*-
\section{Conclusions and Future Work}\label{sec:conclusions}

In this paper we described a mechanism that realizes confidentiality
and anonymity in distributed simulation. The proposed solution has
been implemented in a distributed simulator called AnonSim. AnonSim is
based on the GAIA/ART\`IS simulation framework and the Tor network as
the underlying system to anonymize and make untraceable all the
communications. The proposed approach has been demonstrated as a
viable solution, although at the cost of higher communication
latencies, and a longer initialization phase. However, the benefits
for anonymity and untraceability are evident in many distributed
simulation scenarios.

As a future work, we plan to integrate in AnonSim the fault-tolerance
mechanism based on functional replication described
in~\cite{gda-dsrt-2016}. This would make AnonSim more robust with
respect to spurious failures introduced by the Tor network, while
still supporting anonymity and confidentiality of communications.

\small{
\bibliographystyle{abbrv}
\bibliography{anonsim}

\begin{thebibliography}{10}

\bibitem{okeanos}
Okeanos: Grnet's cloud service.
\newblock https://okeanos.grnet.gr.

\bibitem{HLA}
{IEEE Standard for Modeling and Simulation {(M\&S)} High Level Architecture
  {(HLA)} - Framework and Rules}.
\newblock {\em IEEE Std. 1516-2000}, pages i --22, 2000.

\bibitem{pads}
{PADS: Parallel and Distributed Simulation Research Group}.
\newblock http://pads.cs.unibo.it/, 2018.
\newblock University of Bologna.

\bibitem{tor}
Tor project: anonymity online.
\newblock https://www.torproject.org/, 2018.

\bibitem{EC2}
Amazon.
\newblock Elastic cloud compute.

\bibitem{6266172}
M.~Backes, I.~Goldberg, A.~Kate, and E.~Mohammadi.
\newblock Provably secure and practical onion routing.
\newblock In {\em 2012 IEEE 25th Computer Security Foundations Symposium},
  pages 369--385, June 2012.

\bibitem{Bieber98securityextensions}
P.~Bieber, J.~Cazin, P.~Siron, and G.~Zanon.
\newblock {Security Extensions to ONERA HLA RTI Prototype}.
\newblock In {\em In Proc. of the 1998 Fall Simulation Interoperability
  Workshop}, pages 511--516, 1998.

\bibitem{gda-perf-2005}
L.~Bononi, M.~Bracuto, G.~D'Angelo, and L.~Donatiello.
\newblock Scalable and efficient parallel and distributed simulation of
  complex, dynamic and mobile systems.
\newblock In {\em Proc. of the 2005 Workshop on Techniques, Methodologies and
  Tools for Performance Evaluation of Complex Systems}, Washington, DC, USA,
  2005. IEEE Computer Society.

\bibitem{Chandy:1981:ADS:358598.358613}
K.~M. Chandy and J.~Misra.
\newblock Asynchronous distributed simulation via a sequence of parallel
  computations.
\newblock {\em Commun. ACM}, 24(4):198--206, Apr. 1981.

\bibitem{gda-simpat-2017}
G.~D'Angelo.
\newblock The simulation model partitioning problem: an adaptive solution based
  on self-clustering.
\newblock {\em Simulation Modelling Practice and Theory (SIMPAT)}, 70:1 -- 20,
  2017.

\bibitem{hpcs16}
G.~D'Angelo, S.~Ferretti, and V.~Ghini.
\newblock {Simulation of the Internet of Things}.
\newblock In {\em 2016 International Conference on High Performance Computing
  Simulation (HPCS)}, pages 1--8, July 2016.

\bibitem{gda-simpat-iot}
G.~D'Angelo, S.~Ferretti, and V.~Ghini.
\newblock Multi-level simulation of internet of things on smart territories.
\newblock {\em Simulation Modelling Practice and Theory (SIMPAT)}, 73:3 -- 21,
  2017.

\bibitem{gda-dsrt-2016}
G.~D'Angelo, S.~Ferretti, M.~Marzolla, and L.~Armaroli.
\newblock Fault-tolerant adaptive parallel and distributed simulation.
\newblock In {\em Proc. of the 20th ACM/IEEE International Symposium on
  Distributed Simulation and Real Time Applications (DS-RT)}, pages 37--44,
  Washington, DC, USA, 2016. IEEE Computer Society.

\bibitem{gda-marzolla}
G.~D'Angelo and M.~Marzolla.
\newblock New trends in parallel and distributed simulation: From many-cores to
  cloud computing.
\newblock {\em Simulation Modelling Practice and Theory}, 49:320--335, 2014.

\bibitem{Elkins:2001:SIH:564124.564239}
A.~Elkins, J.~W. Wilson, and D.~Gracanin.
\newblock Security issues in high level architecture based distributed
  simulation.
\newblock In {\em Proc. of the 33Nd Conference on Winter Simulation}, WSC '01,
  pages 818--826, Washington, DC, USA, 2001. IEEE Computer Society.

\bibitem{Fuj00}
R.~M. Fujimoto.
\newblock Parallel discrete event simulation.
\newblock {\em Commun. ACM}, 33(10):30--53, Oct. 1990.

\bibitem{fujimoto}
R.~M. Fujimoto.
\newblock {\em Parallel and distributed simulation systems}, volume 300.
\newblock Wiley New York, 2000.

\bibitem{Fujimoto:2016:RCP:2892241.2866577}
R.~M. Fujimoto.
\newblock Research challenges in parallel and distributed simulation.
\newblock {\em ACM Trans. Model. Comput. Simul.}, 26(4):22:1--22:29, May 2016.

\bibitem{timewarp}
D.~Jefferson.
\newblock {Virtual Time}.
\newblock {\em ACM Transactions Program. Lang. Syst.}, 7(3):404--425, 1985.

\bibitem{1530673}
D.~McGrath, A.~Hunt, and M.~Bates.
\newblock A simple distributed simulation architecture for emergency response
  exercises.
\newblock In {\em Ninth IEEE International Symposium on Distributed Simulation
  and Real-Time Applications}, pages 221--226, Oct 2005.

\bibitem{Mills-Tettey:2003:SIA:786111.786229}
G.~A. Mills-Tettey and L.~F. Wilson.
\newblock Security issues in the abels system for linking distributed
  simulations.
\newblock In {\em Proceedings of the 36th Annual Symposium on Simulation}, ANSS
  '03, pages 135--, Washington, DC, USA, 2003. IEEE Computer Society.

\bibitem{rwp}
M.~Musolesi and C.~Mascolo.
\newblock Mobility models for systems evaluation.
\newblock In {\em Middleware for Network Eccentric and Mobile Applications},
  pages 43--62. Springer Berlin Heidelberg, 2009.

\bibitem{Norling2007SecureDC}
K.~Norling, D.~Broman, P.~Fritzson, A.~Siemers, and D.~Fritzson.
\newblock Secure distributed co-simulation over wide area networks.
\newblock In {\em Proc. of the 48th Scandinavian Conference on Simulation and
  Modeling}, 2007.

\bibitem{oatao2056}
E.~Noulard, J.-Y. Rousselot, and P.~Siron.
\newblock {CERTI, an Open Source RTI, why and how}.
\newblock In {\em Spring Simulation Interoperability Workshop}, pages pp.
  1--11, San Diego, US, 2009.

\bibitem{tcpping}
R.~van~den Berg.
\newblock tcpping: test response times using {TCP SYN} packets.
\newblock http://www.vdberg.org/~richard/tcpping.

\bibitem{5975588}
Z.~H. Zhang, X.~D. Chai, and B.~C. Hou.
\newblock System security approach for web-enabled hla/rti in the cloud
  simulation environment.
\newblock In {\em 2011 6th IEEE Conference on Industrial Electronics and
  Applications}, pages 245--248, June 2011.

\end{thebibliography}
}
\end{document}